\documentclass[conference]{IEEEtran}
\IEEEoverridecommandlockouts
\usepackage{cite}
\usepackage{amsmath,amssymb,amsfonts}
\usepackage{algorithmic}
\usepackage{graphicx}
\usepackage{xcolor}
\usepackage{glossaries}
\usepackage{booktabs}
\usepackage{verbatim}
\usepackage{subcaption}
\usepackage[normalem]{ulem} 
\usepackage{multirow}
\usepackage[super]{nth} 

\newcommand{\HRone}{\mathrm{\mathbf{I}_{HR_1}}}

\newcommand{\LR}{\mathrm{\mathbf{I}_{LR}}}
\newcommand{\LRc}{\mathbf{I}_{\mathrm{LR}_c}}

\newcommand{\Tc}{\mathcal{T}_c}

\newcommand{\HRhatone}{\mathrm{\mathbf{\hat{I}}_{HR_1}}}
\newcommand{\LRone}{\mathrm{\mathbf{I}_{LR_1}}}

\newcommand{\Net}{\mathcal{N}}

\newcommand{\Tinv}{\mathcal{T}^{-1}}

\newcommand{\dcorr}{\mathcal{D}_{\text{con}}}
\newcommand{\duncorr}{\mathcal{D}_{\text{unc}}}
\newcommand{\HRi}{\mathbf{I}_{\mathrm{HR}_{1_i}}}
\newcommand{\LRciset}{ \{\mathbf{I}_{\mathrm{LR}_{c}} \}_i}

\newacronym{voc}{VOC}{Volatile Organic Compound}
\newacronym{bvoc}{BVOC}{Biogenic Volatile Organic Compound}
\newacronym{sr}{SR}{Super-Resolution}
\newacronym{sisr}{SISR}{Single-Image Super-Resolution}
\newacronym{misr}{MISR}{Multi-Image Super-Resolution}
\newacronym{san}{SAN}{Second-order Attention Network}
\newacronym{nlrg}{NLRG}{Non-Locally Enhanced Residual Group}
\newacronym{lsrag}{LSRAG}{Local-Source Residual Attention Groups}
\newacronym{srgan}{SRGAN}{Super-Resolution Generative Adversarial Network}
\newacronym{esrgan}{ESRGAN}{Enhanced SRGAN}
\newacronym{srresnet}{SRResNet}{Super-Resolution ResNet}
\newacronym{msrresnet}{MSRResNet}{Modified SRResNet}
\newacronym{rcan}{RCAN}{Residual Channel Attention Network}
\newacronym{rrdbnet}{RRDBNet}{Residual-in-Residual Dense Block Network}
\newacronym{srcnn}{SRCNN}{Super-Resolution Convolutional Neural Network}
\newacronym{cdf}{CDF}{Cumulative Distribution Function}
\newacronym{hr}{HR}{High Resolution}
\newacronym{lr}{LR}{Low Resolution}
\newacronym{megan}{MEGAN}{Model of Emissions of Gases and Aerosols from Nature}
\newacronym{roc}{ROC}{Receiver Operating Characteristic}
\newacronym{auc}{AUC}{Area Under the Curve}
\newacronym{mse}{MSE}{Mean Squared Error}
\newacronym{nmse}{NMSE}{Normalized Mean Squared Error}
\newacronym{ssim}{SSIM}{Structural Similarity Index Measure}
\newacronym{pcc}{PCC}{Pearson Correlation Coefficient}
\newacronym{lst}{LST}{Land Surface Temperature}
\newacronym{sst}{SST}{Sea Surface Temperature}
\newacronym{vgg}{VGG}{Visual Geometry Group}
\newacronym{pinn}{PINN}{Physics-Informed Neural Networks}
\newacronym{stft}{STFT}{Short Time Fourier Transform}
\newacronym{dft}{DFT}{Discrete Fourier Transform}
\newacronym{dct}{DCT}{Discrete Cosine Transform}
\newacronym{cnn}{CNN}{Convolutional NN}
\newacronym{dl}{DL}{Deep Learning}
\newacronym{nn}{NN}{Neural Network}
\newacronym{mbsr}{MBSR}{Multi-BVOC Super-Resolution}
\newacronym{soa}{SOA}{Secondary Organic Aerosols}
\newacronym{ccn}{CCN}{Cloud Condensation Nuclei}
\newacronym{dwt}{DWT}{Discrete Wavelet Transforms}
\newacronym{ssw}{SSW}{Sea Surface Wind}

\begin{document}
\bstctlcite{IEEEexample:BSTcontrol}

\title{Multi-BVOC Super-Resolution \\ Exploiting Compounds Inter-Connection
\thanks{This work was supported by the Italian Ministry of University and Research (MUR) and the European Union (EU) under the PON/REACT project.}}

\author{\IEEEauthorblockN{
Antonio Giganti,
Sara Mandelli,
Paolo Bestagini,
Marco Marcon,
Stefano Tubaro
}
\IEEEauthorblockA{\textit{Dipartimento di Elettronica, Informazione e Bioingegneria - Politecnico di Milano - Milan, Italy}}
\IEEEauthorblockA{\{antonio.giganti, sara.mandelli, paolo.bestagini, marco.marcon, stefano.tubaro\}@polimi.it}
}

\maketitle

\begin{abstract}
\glspl{bvoc} emitted from the terrestrial ecosystem into the Earth's atmosphere are an important component of atmospheric chemistry. Due to the scarcity of measurement, a reliable enhancement of \glspl{bvoc} emission maps can aid in 
providing denser data for atmospheric chemical, climate, and air quality models.
In this work, we propose a strategy to super-resolve coarse \gls{bvoc} emission maps by simultaneously exploiting the contributions of different compounds. 
To this purpose, we first accurately investigate the spatial inter-connections between several \gls{bvoc} species.
Then, we exploit the found similarities to build a \gls{misr} system, in which a number of emission maps associated with diverse compounds are aggregated to boost \gls{sr} performance. 
We compare different configurations regarding the species and the number of joined \glspl{bvoc}. 
Our experimental results show that incorporating \glspl{bvoc}' relationship into the process can substantially improve the accuracy of the super-resolved maps.
Interestingly, the best results are achieved when we aggregate the emission maps of strongly uncorrelated compounds. This peculiarity seems to confirm what was already guessed for other data-domains, i.e., joined uncorrelated information are more helpful than correlated ones to boost \gls{misr} performance. 
Nonetheless, the proposed work represents the first attempt in \gls{sr} of \gls{bvoc} emissions through the fusion of multiple different compounds.
\end{abstract}


\section{Introduction}
\label{sec:introduction}

Many chemicals are produced by terrestrial ecosystems, including volatile or semi-volatile compounds, and released into the atmosphere. Among all these chemicals, \glspl{bvoc} have been recognized as significant contributors to air quality and climate change due to their large emission amount and high reactivity~\cite{bvoc_risk_2023, guenther_model_2012, calfapietra_role_2013, penuelas_bvocs_2010}. 
For instance, \glspl{bvoc} are important precursors of ozone and secondary organic aerosols, which can negatively impact human health and vegetation growth~\cite{bvoc_exchange_2021, opacka_isoprene_2021}. 

To assess air quality and climate conditions, it is crucial to accurately estimate the amount of \gls{bvoc} emissions, both historically and in the present and future~\cite{cai_bvoc_scientometric_2021, guenther_model_2012, hewitt_bvoc_quantification_2011}. 
Although various ground-based techniques are available to sample \gls{bvoc} emissions across different scales
~\cite{opacka_isoprene_2021, hewitt_bvoc_quantification_2011}, the available measurements are limited in space and time, making them less suitable for reliably simulating atmospheric, climate, and forecasting models.

To tackle the lack of measurements, we propose to increase the spatial resolution of \gls{bvoc} emissions by refining a coarser grid. 
This problem falls into the general category of image \gls{sr} tasks, which aim at enhancing the pixel resolution of a digital image.
The goal is to 
find a 
suitable mapping between the \gls{lr} image at hand and its corresponding \gls{hr} version in a way that ensures high-quality upscaling. 

Plenty of work has been done to enhance classic $8$-bit imagery like photographs or medical imaging. 
Several approaches
have been proposed, and \gls{dl}-based methods have shown to outperform classical methods~\cite{wang_review_2022} thanks to their ability to learn 
spatial
features from huge 
datasets~\cite{medina_dl_comparison_sr_2020}.
However, \gls{sr} of 2D data linked to physical measures (like \gls{bvoc} emissions) is less explored and not straightforward. Data involving physical acquisitions are always connected with a meaningful measurement unit; moreover, they might report a sparse nature, numerous outliers, and wide dynamic ranges. 
Therefore, the modification of standard \gls{sr} techniques is usually required \cite{geiss_chem_sim_sr_2022, giganti2023enhancing}. 


The majority of the proposed works mainly exploit information coming from a single observation~\cite{brecht_sisr_wind_2022, nguyen_sisr_lst_2022, yasuda_micrometeorology_2022, stengel_solar_wind_sr_2020, giganti2023enhancing}, i.e., the tackled problem can be formulated as a \gls{sisr} task. 
Nonetheless, combining information from multiple different observations belonging to the same or similar domains proved a useful strategy to increase \gls{sr} performance~\cite{tarasiewicz_misr_gnn_2021, tr_misr_2022}. This converts into a \gls{misr} task.
For example, authors in~\cite{lloyd_misr_optically_2022} focused on improving the resolution of sea surface temperature by exploiting both optical and thermal images. 
A combination of information related to sea surface temperature, sea surface wind, and other remote-sensing products was proposed in~\cite{salinity_2022} to reconstruct a denser ocean subsurface salinity. Authors of~\cite{liu_air_pollution_2019} exploited a pollution field's external factors and spatial-temporal dependencies to increase its resolution.  In~\cite{wang_precipitation_sr_2021}, \gls{sr} of precipitation data was performed leveraging \gls{hr} topography maps and various \gls{lr} maps of sea level pressure and air temperature.

In this paper, we propose to fuse the emission maps of different \glspl{bvoc} to provide denser emission maps. In particular, multiple \gls{lr} emission maps are combined together and processed through a \gls{nn} to super-resolve emissions of a specific compound. 
To do so, we thoroughly explore the inter-connections between several \gls{bvoc} species, investigating their structural similarities and spatial correlations. 
In our experiments, we compare various configurations regarding the species and number of aggregated compounds used for facing the proposed \gls{misr} objective. 
We show that a specific selection of \gls{bvoc} species based on their related inter-connection can provide a performance gain in \gls{sr} of emission maps. Interestingly, the aggregation of less correlated \glspl{bvoc} proves more beneficial than the joint contribution of related compounds, similar to what was guessed in \cite{lloyd_misr_optically_2022} for different domain data.

\section{BVOCs Inter-Connection}
\label{sec:correlation}

The emissions related to different \glspl{bvoc} are often inter-related, and changes in the emission rate of one compound can affect the emission rate of others. 
Several studies explored the complex underlying mechanism which governs these correlations~\cite{cai_bvoc_scientometric_2021, stavrakou_top-down_consistency_2015}. 
In principle, \gls{bvoc} emissions strongly vary depending on the species of vegetation and on environmental, meteorological, and physiological factors. Moreover, environmental stress phenomena can induce other diving factors~\cite{midzi_stress_induced_2022, feldner_abiotic_2022}.

In this section, we investigate the spatial inter-connection of \gls{bvoc} emission maps related to multiple compounds.
To do so, we select \gls{bvoc} emission maps from the most recent global coverage biogenic emission inventory~\cite{sindelarova_high-resolution_2022}. This inventory provides emissions from $25$ different compounds, including isoprene, monoterpene, sesquiterpene, methanol, and other main \gls{bvoc} species. 

Given a reference biogenic compound, we explore its spatial inter-connection with other compounds by computing the \gls{pcc} and the \gls{ssim} between their emission maps. We always compare emission maps related to the same geographical area and acquisition date to obtain reasonable comparisons. 

\begin{figure}[t]
\centering
\includegraphics[width=.8\columnwidth]{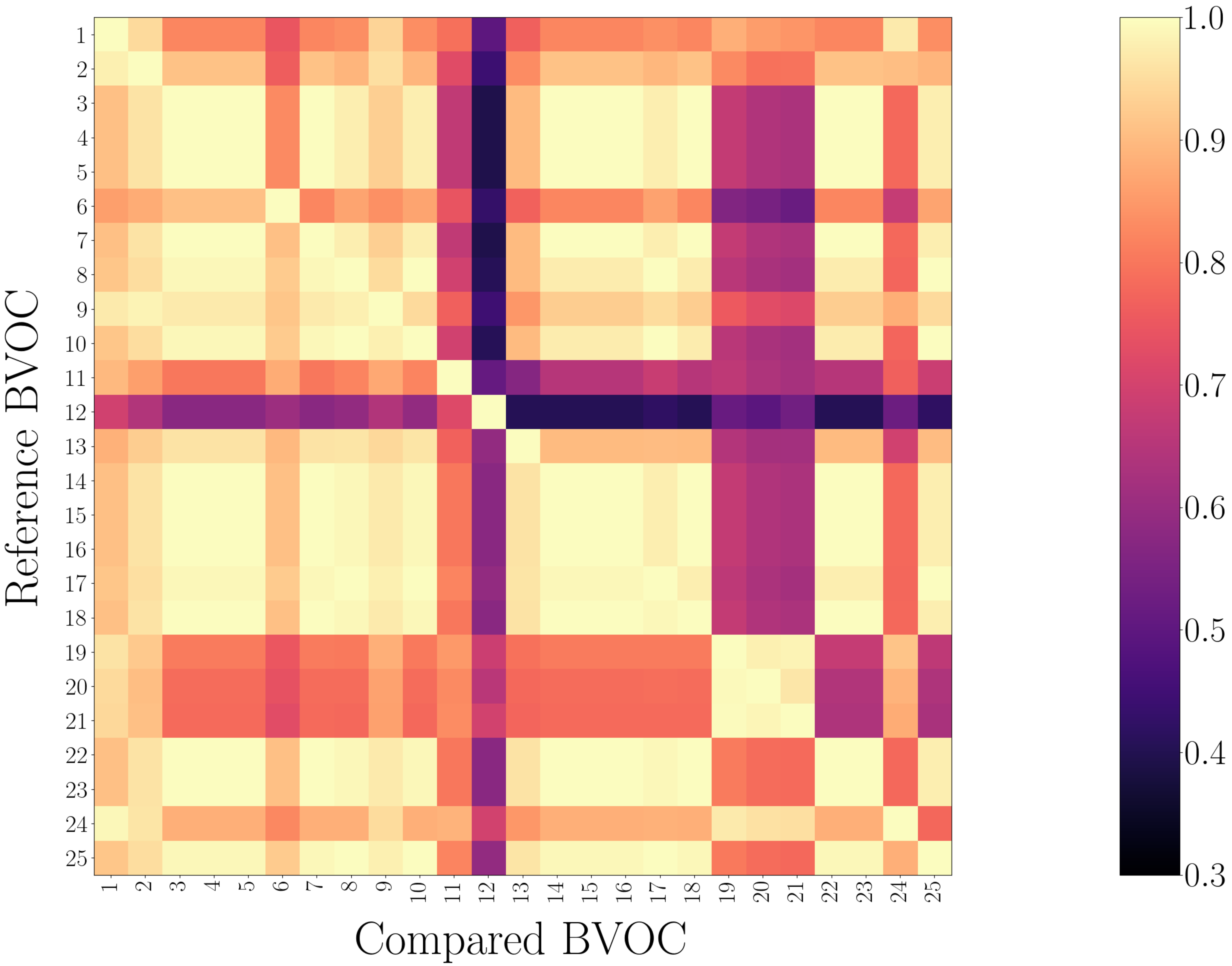}
\caption{Spatial inter-connection of multiple \gls{bvoc} emission maps.}
\label{fig:inter-correlations}
\vspace{-15pt}
\end{figure}

Fig.~\ref{fig:inter-correlations} depicts the computed inter-connection measures. Given the intrinsic commutative property of \gls{ssim} and \gls{pcc}, the upper triangular region reports the \gls{ssim}, and the lower triangular region reports the \gls{pcc}. 
The elements along the diagonal always equal $1$ for both metrics, since the reference and the compared compounds are the same.

It is worth noticing that \gls{pcc} and \gls{ssim} are in accordance one another, i.e., high \gls{pcc} corresponds to high \gls{ssim} as well, and vice versa.
Emission maps of different compounds 
actually contain some inter-correlations, i.e., the 
inter-connection matrix does not present a perfect diagonal behavior.

These preliminary investigations motivate our proposed approach, that is, the joint exploitation of different \gls{bvoc} emission maps for super-resolving \gls{lr} maps. 
Following prior works in \gls{misr} over satellite-derived physical measures~\cite{lloyd_misr_optically_2022, salinity_2022, wang_precipitation_sr_2021}, we believe that the related cross-information among different compounds enables to boost performance with respect to the simple \gls{sisr} scenario.
In the next section, we present the proposed \gls{mbsr} strategy which is built upon this intuition.

\section{Multi-BVOC Super Resolution}
\label{sec:multi_bvoc}

\glsreset{lr}
\glsreset{hr}
\glsreset{sisr}
\glsreset{misr}
\glsreset{mbsr}

\begin{figure}[t]
\centering
    \begin{subfigure}[a]{0.31\textwidth}
        \centering
        \includegraphics[width=\textwidth]{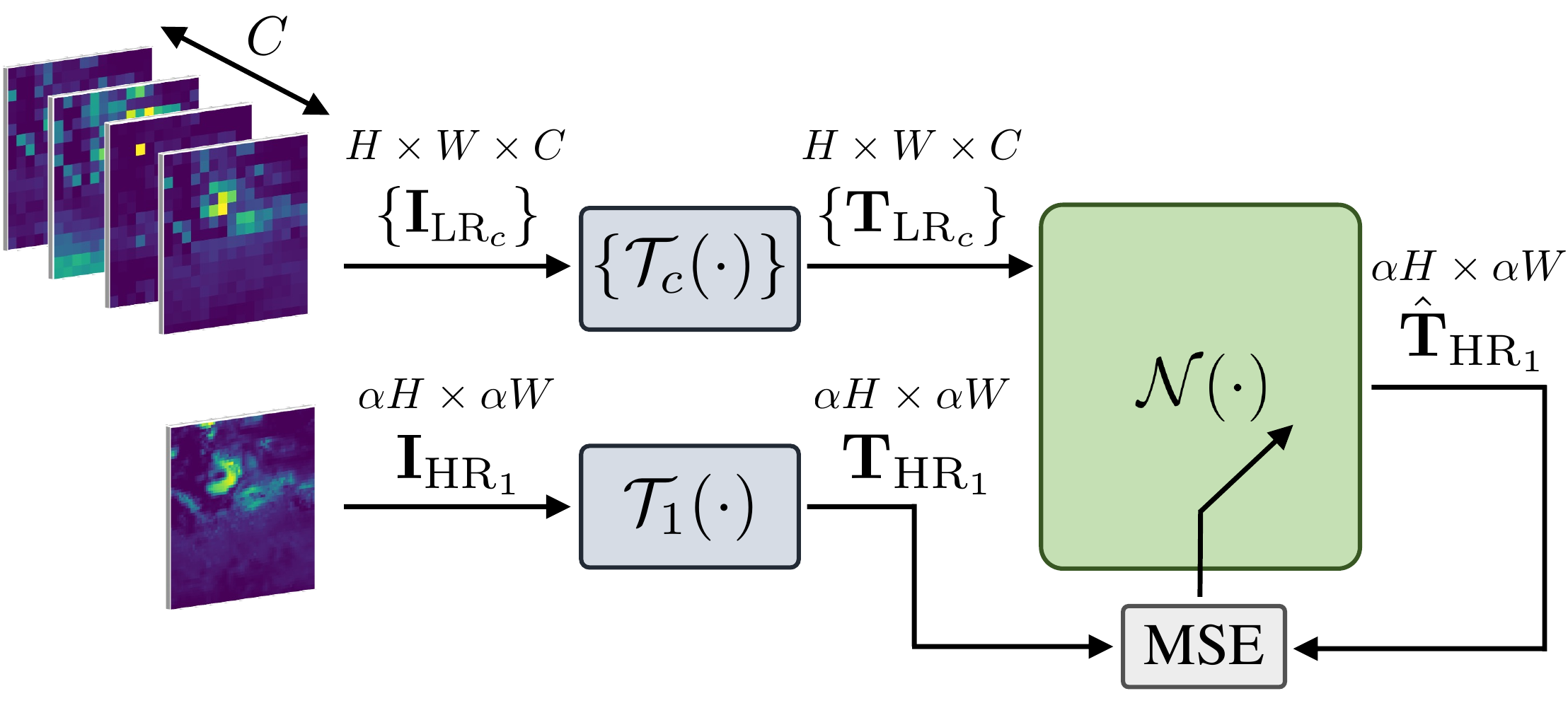}
        \caption{Training.}        
        \label{fig:misr training}
    \end{subfigure}
    \begin{subfigure}[b]{\columnwidth}
        \centering
        \includegraphics[width=\textwidth]{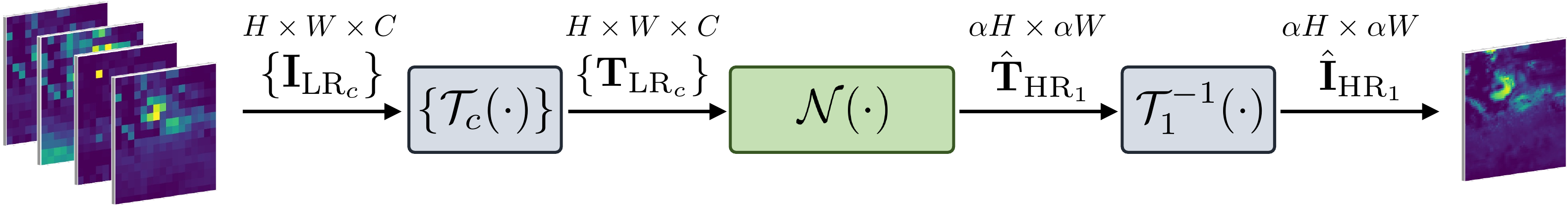}
        \caption{Deployment.}
        \label{fig:misr deployment}
    \end{subfigure}
\caption{The proposed \gls{mbsr} method: (a) training and (b) deployment phase.}   
\label{fig:misr_system}
  \vspace{-15pt}
\end{figure}

The proposed work for \gls{mbsr} aims at exploiting spatial inter-connections between different \glspl{bvoc} to super-resolve \gls{lr} emissions.
Differently from \gls{sisr} tasks, we tackle a \gls{misr} problem, in which a \gls{hr} emission map of a reference \gls{bvoc} is recovered by exploiting multiple \gls{lr} emission maps of different \glspl{bvoc}. 

Following the considerations done in Section~\ref{sec:correlation}, we select $C$ compounds according to their spatial inter-connections, stacking their emission maps in a set defined as $\{ \LRc \}, c = 1, ..., C$. 
The set includes the reference \gls{bvoc} \gls{lr} emission $\LR_1$ associated with the index $c=1$ and other joined compounds' \gls{lr} emissions $\{ \LRc \}, c = 2, ..., C$. 
The reference compound emission $\LR_1$ is the one we want to super-resolve.
We propose to estimate an \gls{hr} emission map as
\begin{equation}
    \hat{\mathbf{{I}}}_{\mathrm{HR}_1} = \Tinv_1(\Net( \{ \Tc(\LRc) \})), 
\end{equation}
where $\{\Tc(\cdot) \}$ collects $C$ different data transformations $\Tc(\cdot)$, each one applied to the \gls{lr} emission $\LRc$ of the $c\text{-th}$ compound, $\Net(\cdot)$ is a \gls{nn} operator, and $\Tinv_1(\cdot)$ is the inverse transformation related to the reference \gls{bvoc}, with $c=1$.

Fig.~\ref{fig:misr_system} depicts a sketch of the proposed \gls{mbsr} methodology.
We model $\{\LRc\}$ and $\{ \Tc(\LRc) \}$ as tensors with size of $H\times W \times C$, where $H$ and $W$ represent height and width of the emission maps and $C$ the number of compounds adopted for performing \gls{mbsr}. 
Matrices ${\mathbf{{I}}}_{\mathrm{HR}_1}$ and its estimation $\hat{\mathbf{{I}}}_{\mathrm{HR}_1}$ have size $\alpha M\times \alpha N$, with $\alpha > 1$ indicating the super-resolution factor (i.e., how much we increase the resolution). 
The training phase involves ($\{ \LRc \}$, $\mathbf{{I}}_{\mathrm{HR}_1}$) emissions as input (see Fig.~\ref{fig:misr training}).
At testing stage (see Fig.~\ref{fig:misr deployment}), given a set of $C$ \gls{lr} emissions $\{ \LRc \}$, we estimate $\hat{\mathbf{{I}}}_{\mathrm{HR}_1}$. 

In our past investigations~\cite{giganti2023enhancing}, we showed that a suitable data transformation is required to deal with \gls{bvoc} emissions since they are characterized by sparsity, extremely small values, and wide dynamic ranges (from $10^{-30}$ to $10^{-9}$ [kg/m\textsuperscript{2}s]).
As a matter of fact, \gls{bvoc} emissions can present many outliers due to the large spatial diversity of the environmental factors driving the emission process, such as meteorology, type of vegetation, seasonal cycle, and atmospheric composition~\cite{sindelarova_high-resolution_2022, giganti2023enhancing}. 
For this reason, we adopt a set $\{ \Tc(\cdot) \}$ of non-parametric transformations that force emission values to follow a uniform distribution between $0$ and $1$~\cite{scikit-learn, giganti2023enhancing}. Notice that these transformations are compound-specific, i.e., every $\Tc(\cdot)$ 
strictly depends on the $c$-th compound since it is based on statistical information extracted from its emissions. 

It is worth noticing that the choice of a \gls{nn} as \gls{sr} operator is not only motivated by the superior performances of novel \gls{dl}-based methods with respect to classical approaches. 
In a \gls{misr} context, the advantage of solutions leveraging \gls{dl} is that these do not require to discover a formal relationship between the different stacked \gls{bvoc} emission maps.
The training process benefits from any advantageous connections existing between multiple distinct types of input images~\cite{lloyd_misr_optically_2022}.
In particular, we exploit the \gls{san} architecture~\cite{dai_sansisr_2019}, which 
can be considered the state-of-the-art in this field~\cite{giganti2023enhancing}.

\section{Experimental Analysis}
\label{sec:experiments}

\subsection{Dataset Collection}
\label{ssec:dataset}
To perform our investigations, we use the most up-to-date and highest-resolution global coverage biogenic emission inventory available in the literature~\cite{sindelarova_high-resolution_2022}. 
This inventory provides emissions from various \glspl{bvoc}, covering the entire Earth's surface for the period of 2000-2020 at a high spatial resolution of $0.25^{\circ}\times0.25^{\circ}$, which is approximately $28$km$\times28$km for each cell in continental regions. 

As suggested in \cite{giganti2023enhancing}, we slice each emission map, which has a grid of $1440\times720$ cells, into non-overlapping patches of $64\times64$ cells. This step makes the computations more manageable and enables to assume minimal radial distortions due to the Earth's curvature. 
For every \gls{bvoc}, we end up with $81957$ distinct patches that we consider as \gls{hr} references.
The associated \gls{lr} patches are generated by performing bicubic downsampling, resulting in $16\times 16$ emission maps. 
Thus, we aim at super-resolving \gls{lr} emission maps with $1^{\circ}\times1^{\circ}$ spatial resolution into \gls{hr} emission maps with $0.25^{\circ}\times0.25^{\circ}$ resolution, which corresponds to a scale factor of $\alpha=4$.
We adopt this factor since it is more challenging with respect to smaller ones, as we found in our previous investigations~\cite{giganti2023enhancing}.

Isoprene is the most prevalent and impactful \gls{bvoc} in the inventory, accounting for approximately half of the total \gls{bvoc} emissions~\cite{calfapietra_role_2013}.
Based on this, we focus on the \gls{sr} of isoprene emission maps.
To solve the \gls{mbsr} task, 
we create $ \{\LRc \} $ tensors 
by stacking one \gls{lr} patch of isoprene and \gls{lr} patches related to the same geographical area but to different \glspl{bvoc}. 
Our final dataset is defined as $\mathcal{D} = \{ \LRciset, \HRi \}$, for $c = 1, ..., C$ and $i = 1, ..., 81957$.
To select the joined compounds, we exploit the inter-connection metrics shown in Section~\ref{sec:correlation}.
In Section~\ref{sec:results}, we provide more details on the selected compounds, comparing different experimental scenarios.

\subsection{Training Setup}
\label{ssec:training}
As reported in \cite{giganti2023enhancing}, each compound needs a different data transformation based on statistical information derived from its \gls{hr} emissions. To estimate the transformation set $\{ \Tc(\cdot) \}, c = 1, ..., C$, we follow the experimental setup shown in \cite{giganti2023enhancing}.

We divide our dataset into train, validation, and test sets with $70/20/10$ percentage amount, respectively.
We combine the ADAM optimizer with the Cosine Annealing learning rate scheduler~\cite{loshchilov_sgdr_2017},
setting the initial learning rate to $10^{-4}$, with a minimum value of $10^{-7}$. 
We set a maximum number of $3\cdot 10^{5}$ iterations, validating the model each $10^{3}$ iterations and
stopping the algorithm if the validation loss does not further improve after $10$ validation steps.

\section{Experimental Results}
\label{sec:results}


\subsection{BVOC Inter-Connection Analysis}
\label{ssec:compounds inter correlation}

\begin{figure}[t]
\centering
\includegraphics[width=\columnwidth]{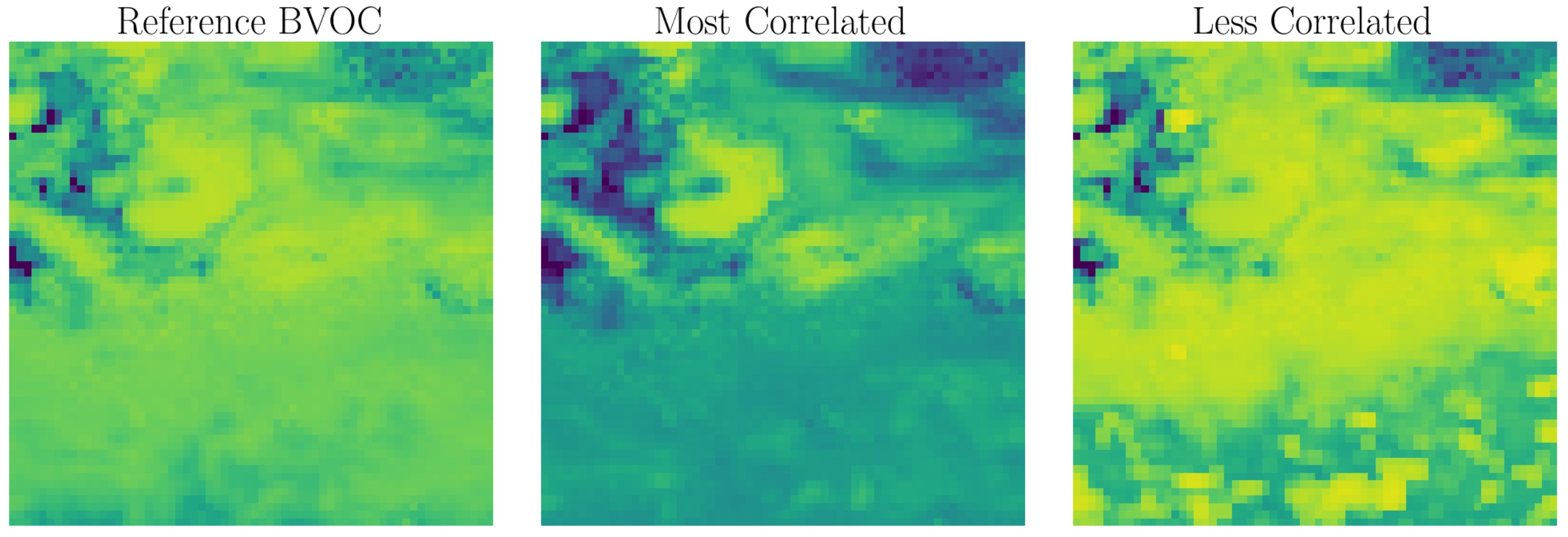}
\caption{Example of most correlated and uncorrelated \glspl{bvoc} with respect to isoprene (reference \gls{bvoc}). 
}
\label{fig:corr_bvoc_example}
\vspace{-15pt}
\end{figure}

\begin{figure*}[t!]
\centering
\includegraphics[width=0.9\textwidth]{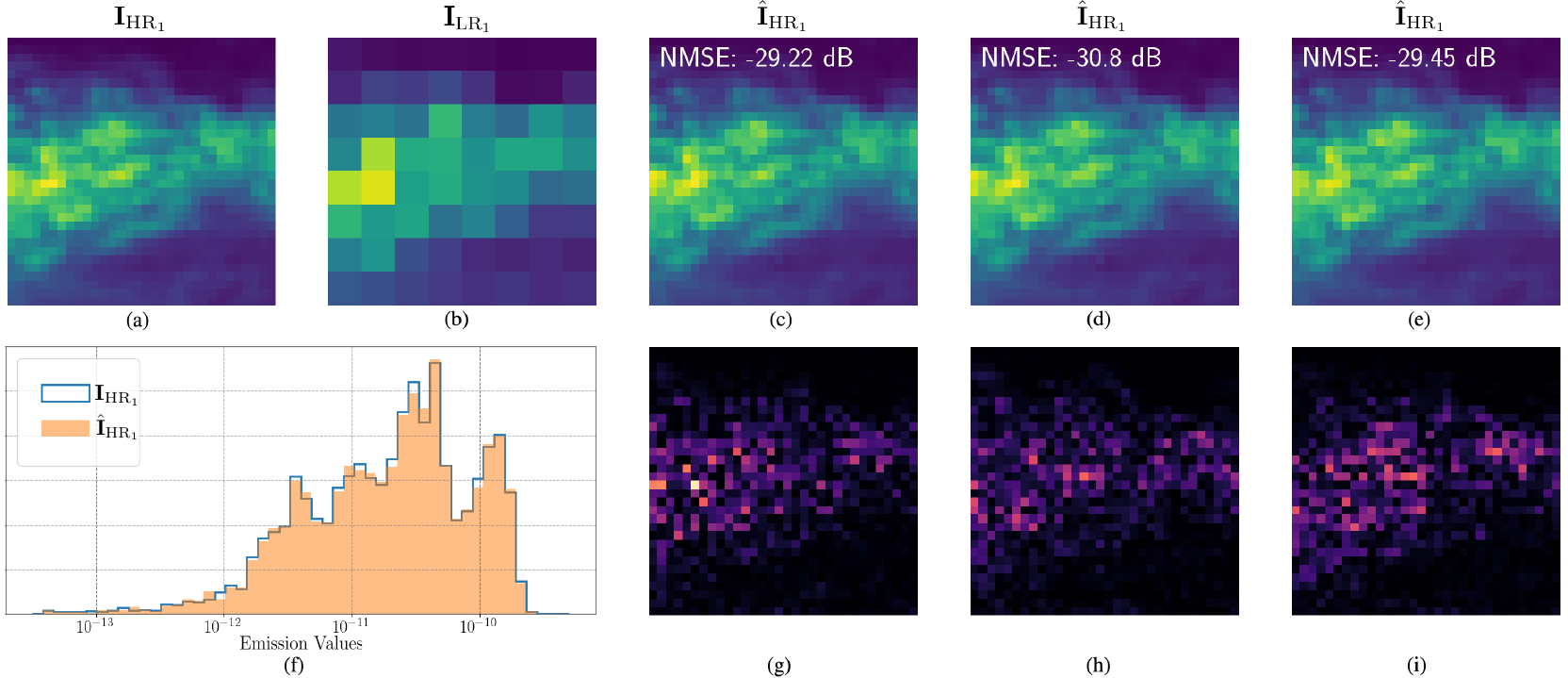}
\caption{Example of emission \gls{sr}. 
(a) is the original \gls{hr} isoprene emission map $\HRone$; (b) shows its related \gls{lr} version $\LRone$; (c), (d), (e) are the \gls{sr} emissions $\HRhatone$ for the cases $C=2$, $C=3$ (best case) and $C=4$, respectively; (g), (h), (i) their relative absolute error, computed as $|\HRone - \HRhatone|$ (the brighter, the higher); (f) reports the histogram of the emission values of (a) ($\HRone$) compared with the best super-resolved emission map in (d), ($\HRhatone$).}

\label{fig:sr performance}
\vspace{-10pt}
\end{figure*}


Our main goal is super-resolving isoprene emission maps, corresponding to index $11$ in the \gls{bvoc} inter-connection matrix shown in Fig.~\ref{fig:inter-correlations}. 
Due to the substantial accordance between \gls{pcc} and \gls{ssim}, we select the $3$ most similar and the $3$ least similar compounds based on their related \gls{ssim}. 

From higher to lower, acetaldehyde, sesquiterpenes and, formaldehyde are the $3$ best correlating compounds with isoprene. 
On the contrary, 
the 3 less correlated ones are
MBO (2-methyl-3-buten-2-ol), methanol, and $\beta$-pinene.
Fig.~\ref{fig:corr_bvoc_example} shows an example of the spatial coherence between the reference compound (i.e., isoprene) and the most and least correlated compounds.

Vegetation emits all the selected compounds, and some are also partly products of the isoprene's oxidation in the atmosphere.
In line with past works~\cite{stavrakou_top-down_consistency_2015, karl_atmo_cleaning_vegetation_2010}, we found that the most inter-connected compounds are those influenced by similar environmental factors since similar plant species produce them. 
On the contrary, the compounds exhibiting a fairly low inter-connection are mainly produced by isoprene's oxidation~\cite{wolfe_form_iso_2016}.
We believe that these results are corroborated by the fact that oxidation phenomena are closely linked to atmospheric chemistry and meteorology, thus weakly related to geographic topology such as vegetation type.

\subsection{Compound Selection}
\label{ssec:channel selection}

\begin{table}[t]
\centering
\caption{\gls{sr} results for connected ($\dcorr$) and unconnected ($\duncorr$) compounds. 
For $C=2$, \nth{1} and \nth{2} configurations denote the pairs (isoprene, acetaldehyde) and (isoprene, sesquiterpene)
for $\dcorr$, and 
the pairs (isoprene, MBO) and (isoprene, methanol)
for $\duncorr$, respectively. In bold, the best achieved results.}
\label{tab:misr}
\resizebox{\columnwidth}{!}{%
\begin{tabular}{llrccc}
\toprule
 &  &  & $C=2$ & $C=3$ & $C=4$ \\ 
Dataset &  &  & SSIM / NMSE {[}dB{]} & SSIM / NMSE {[}dB{]} & \multicolumn{1}{l}{SSIM / NMSE {[}dB{]}} 
\\ \midrule 
\multirow{2}{*}{$\dcorr$} &  & \nth{1} & $0.988$ / $-21.38$ & \multirow{2}{*}{$0.987$ / $-20.45$} & \multirow{2}{*}{$0.986$ / $-20.79$} \\ 
 &  & \nth{2} & $0.986$ / $-20.92$ &  &  \\ 
 \midrule 
\multirow{2}{*}{$\duncorr$} &  & \nth{1} & $0.988$ / $-21.09$ & \multirow{2}{*}{$\mathbf{0.989}$ / $\mathbf{-21.93}$} & \multirow{2}{*}{$0.986$ / $-20.51$} \\ 
 &  & \nth{2} & $0.987$ / $-21.31$ &  &  \\ \bottomrule
\end{tabular}%
}
  \vspace{-10pt}
\end{table}

In this section, we investigate how much leveraging the inter-connection between compounds benefits the \gls{sr} process. 
To this purpose, we create different datasets based on the compound inter-connections, stacking multiple emission maps along the channel dimension of the tensors, as explained in~\ref{ssec:dataset}.
For brevity's sake, we reduce the potential compounds' combinations by selecting only strongly connected or poorly connected ones. 
The datasets of compounds that show a strong inter-connection with isoprene are generally denoted as $ \dcorr$, while $\duncorr$ indicates the datasets including less connected compounds. 

For both $\dcorr$ and $\duncorr$ categories, we consider the scenarios in which \gls{lr} emission maps of $2, 3$, or $4$ different compounds are aggregated to estimate \gls{hr} emission maps of isoprene. 
For the case $C=2$ (i.e., when $2$ compounds are joined), we investigate two different configurations that correspond to joining isoprene with either acetaldehyde or sesquiterpenes in the $\duncorr$ scenario, and to joining isoprene with either MBO or methanol in the $\duncorr$ scenario.
The case $C = 3$ aggregates to isoprene both acetaldehyde and sesquiterpenes in the $\duncorr$ scenario and both MBO and methanol in the $\duncorr$ scenario. 
Finally, $C = 4$ considers the $3$ most correlating \glspl{bvoc} for $\dcorr$ and the $3$ least correlating \glspl{bvoc} for $\duncorr$.

Table~\ref{tab:misr} shows the achieved results in terms of average \gls{ssim} and \gls{nmse}, where \gls{nmse} is defined as the MSE computed between $\mathbf{{I}}_{\mathrm{HR}_1}$ and $\hat{\mathbf{{I}}}_{\mathrm{HR}_1}$, normalized by the average of ${\mathbf{{I}}^2_{\mathrm{HR}_1}}$. 

All configurations return good results. 
There are no remarkable differences between the configurations related to $C = 2$, i.e., there is not a specific compound which reveals more advantageous when paired with isoprene. 
On average, the configurations associated with $C = 4$ report the worst results. This might be due to some difficulties encountered by the training process in handling too many different data, thus compromising convergence. 
On the other hand, the proposed \gls{mbsr} achieves a remarkable boost in using two additional uncorrelated compounds. 
In specific, \gls{sr} benefits from information extracted from isoprene and the two mostly uncorrelated compounds, i.e., when $C = 3$, dataset $\duncorr$.

We believe 
the spatial difference in the emission patterns favors extracting a distinctive feature by the \gls{nn}, enabling to enhance the \gls{sr} performance.
Similar considerations were reported in~\cite{lloyd_misr_optically_2022}, where the authors found that adding complementary information, even though less correlated with the data to be super-resolved, proved helpful for improving the \gls{sr} results. 
Contrarily, the authors noticed that a positive correlation 
might result in 
redundancy which does not give an improvement in performance, and our achieved results seem to confirm this behaviour. 


Fig.~\ref{fig:sr performance} shows examples of the proposed \gls{mbsr} considering different configurations. 
In particular, we depict results for $\duncorr$ in scenarios $C = 2, 3, 4$ (see Figs.~\ref{fig:sr performance}c-e).
To enhance the details, the emission maps correspond to 
cropped versions of the original emission patches. At the same time, \glspl{nmse} are referred to the entire images, to be comparable with the results in Table~\ref{tab:misr}.

It is worth noticing that the best configuration of Table ~\ref{tab:misr}, i.e., $\duncorr$ and $C=3$, guarantees excellent results, especially in areas of high emission values. 
This can be seen by comparing the absolute reconstruction errors associated with diverse configurations, computed as $|\HRone - \HRhatone|$ (see Figs.~\ref{fig:sr performance}g-i).
In Figs.~\ref{fig:sr performance}g and ~\ref{fig:sr performance}i, 
we can observe that higher deviations are reached at the areas with greater emission of the compound. 
This behavior is much more limited for the best configuration result shown in Fig.~\ref{fig:sr performance}h.
For a more detailed comparison, Fig.~\ref{fig:sr performance}f compares the histograms of the ground-truth and of the estimated emission map through the best configuration, highlighting the significant reconstruction quality of the proposed methodology.

\section{Conclusions}
\label{sec:conclusions}

In this paper, we proposed the \glsfirst{mbsr} method,
which combines multiple biogenic compounds to super-resolve coarse emission maps. 
To the best of our knowledge, there are no prior works investigating \glsfirst{misr} problems on \gls{bvoc}. Nonetheless, \gls{sr} of \gls{bvoc} emissions is paramount to fill the lack of measurements for reliable atmospheric, climate, and forecasting models simulations.

To determine which and how many \glspl{bvoc} should be combined, we conducted careful investigations on the spatial inter-connection between the most abundant \glspl{bvoc} in nature.
Then, we leveraged these inter-connections to drive the \gls{sr} process. We experimented various configurations, aggregating the emission maps of different number of \glspl{bvoc} with high and low spatial correlation.
Interestingly, we found that joining poorly correlated compounds can effectively boost the \gls{sr} performance, preserving spatial patterns and fine-scale structures.


Future works will concern 
the comparison of our proposed strategy with the state-of-the-art \gls{sr} techniques for multispectral images, in which several related images are super-resolved simultaneously.

In addition, we will investigate the adoption of \gls{sr} methods that embed known properties of the underlying physical system behind the \gls{sr} process, and enforcing strict physical agreement and consistency~\cite{geiss_chem_sim_sr_2022} between \gls{lr} and super-resolved data.



\bibliographystyle{IEEEtran}
\bibliography{ref.bib}

\end{document}